\documentclass[JCAP,nofootinbib,superscriptaddress]{revtex4-1}
\input epsf
\usepackage{graphics}
\usepackage{amsmath,amssymb} 
\usepackage{color} 
\usepackage{dcolumn}
\usepackage{hyphenat}
\usepackage{multirow}

\usepackage{graphicx}

\def\be{\begin{equation}}
\def\ee{\end{equation}}
\def\ba{\begin{eqnarray}}
\def\ea{\end{eqnarray}}

\newcommand{\beqa}{\begin{eqnarray}}
\newcommand{\eeqa}{\end{eqnarray}}

\newcommand{\beq}{\begin{equation}}
\newcommand{\eeq}{\end{equation}}

\newlength{\tskip}
\newlength{\colwidth}

\begin{document}

\title{Primordial black hole formation during cosmic phase transitions}

\author{Karsten Jedamzik} 
\affiliation{Laboratoire de Univers et Particules de Montpellier, UMR5299-CNRS, Universite de Montpellier, 34095 Montpellier, France}

\begin{abstract}
Primordial black holes (PBHs) may be part of the dark matter. 
It is shown here that PBHs form more easily during cosmic phase transitions. 
For approximately scale-invariant primordial curvature fluctuations the PBH mass function may therefore leave a record of the thermal history of the early Universe. In particular a peak is expected on the $1.9 M_{\odot}$ scale due to
the cosmic QCD transition.   
\end{abstract}

\maketitle

\subsection{General Considerations}

Primordial black hole (PBH) formation during cosmic phase transitions may
be particularly efficient. In what follows we only discuss the case of
PBH formation due to pre-existing curvature fluctuations produced during inflation. We omit a discussion of other scenarios 
\cite{Crawford:1982yz,Hall:1989hr} where substantial sub-horizon size density fluctuations are generated due to the dynamics of the
phase transition itself. Consider a pre-existing slightly non-linear density
fluctuation when it crosses into the horizon, i.e. of size $r\approx r_H$,
where $r_H = 2 c t = c/H$, is the horizon length during a radiation dominated Universe with $c$ the speed of light, $t$ cosmic time, and $H$ the Hubble expansion rate. Approximate the overdense
region to be in vacuum, i.e. not having cosmic matter around it. One can then
compute the Schwarzschild radius of the overdense region
\begin{equation}
r_S = \frac{2GM}{c^2} = \frac{2G}{c^2}\frac{4\pi}{3}\rho r_H^3 = \frac{2G}{c^2}\frac{4\pi}{3}\rho \biggl(
\frac{c}{H}\biggr)^3 = \frac{2G}{c^2} \frac{c^3}{2 H G} = \frac{c}{H} = r_H\, ,
\end{equation}
where in the second equality we have used the horizon mass and in the fourth
equality the Friedman equation $H^2 = 8\pi G\rho /3$. Here $G$ and $\rho$ are
gravitational constant and energy density, respectively. It is seen that the overdense
matter contained on the horizon scale is close to its own Schwarzschild radius
and therefore close to forming a PBH. It is known that the overdensity of a sub-horizon region does not further grow in a radiation dominated Universe , such that a PBH
will either form at horizon crossing, or never, since $r_S < r_H$ during the subsequent evolution. What is the role of pressure forces during PBH 
formation ? From astrophysics we are used to the concept that an overdense
region may only gravitationally collapse if its size is large enough, larger than the Jeans length. This
is the case if the pressure response time, i.e. the time for a pressure (sound) wave to travel over the entire overdense region, $t_p = r/c_s$, with $c_s$ the
speed of sound, is larger than the gravitational collapse time scale, 
$t_{g}\approx 1/\sqrt{G\rho}$. Using the speed of sound $c_s= c/\sqrt{3}$
of a relativistic gas, as appropriate in a radiation dominated Universe, one
may easily derive the Jeans length
\begin{equation} 
r_J\approx c_s/\sqrt{G\rho} = \frac{\sqrt{8\pi}}{3}\frac{c}{H}\approx 1.67 r_H
\end{equation}
It is seen that forces due to pressure gradients are important. The PBH formation process is thus a competition between gravity and pressure forces.
This implies that during any epochs in the early Universe where pressure forces
are reduced due to a varying equation of state, PBHs may form more easily than
during a purely radiation dominated Universe. As PBH formation is a rare 
process, occuring on an exponential tail (cf. Chapter 3), the slightest reduction of the overdensity required to form a PBH, may lead to a very significant enhancement of PBH formation. For an approximately scale-invariant spectrum
of pre-existing inflationary curvature perturbations, the PBH mass function
should therefore be peaked at the approximate horizon masses during epochs
where a softening of the equations of state occurs. 

\subsection{Generic First-Order phase Transitions}

Consider a generic first order phase transition. During such a transition
there is low-energy phase and a high-energy phase of the matter. These phases
may coexist at coexistence temperature $T_c$. In thermodynamic equilibrium
the quantities $p = -(\partial E/\partial V)_S$ and $T = (\partial E/\partial S)_V$,
which define pressure $p$ and temperature $T$, are continuous at temperature $T_c$. Here $E$, $V$, and $S$ are energy, volume, and entropy, respectively.
During the phase transition, when the system "cools", high-energy phase with
energy density $\rho_h = \rho_l + L$ is converted to low-energy phase, while 
the system stays at the same temperature $T_c$. Here $L$ is latent heat.
Only when all high-energy phase is converted to low-energy phase, will the
temperature drop below $T_c$. Borrowing the parametrisation of the MIT model
(developed to describe a first order QCD transition) one may
write
\begin{eqnarray}
p_l(T) = \frac{1}{3}g_l \frac{\pi^2}{30} T^4\quad ; \quad p_h(T) = \frac{1}{3}g_h \frac{\pi^2}{30} T^4 - B\\
\rho_l(T) = g_l \frac{\pi^2}{30} T^4\quad ; \quad \rho_h(T) = g_h \frac{\pi^2}{30} T^4 + B \label{eq:4}\\
s_l(T) = \frac{4}{3}g_l \frac{\pi^2}{30} T^3\quad ;\quad  s_h(T) = \frac{4}{3}g_h \frac{\pi^2}{30} T^3\quad . \label{eq:5}
\end{eqnarray}
In the above $s$ is entropy density and $g$ are statistical weights
(i.e. essentially the number of relativistic species in a phase). Furthermore
the "bag constant" $B$ acts essentially as a vacuum energy density for the
high energy phase. In the QCD case it was introduced to account for the
interaction energy associated within the strongly interacting quark-gluon plasma. 
During the middle of a phase transition the plasma consists of bubbles of
one phase surrounded by the other phase. Over suffienctly large regions,
i.e. $l >> r_b$, where $r_b$ is the typical size of a bubble, one
may define an average density and entropy
\begin{eqnarray}
\langle p \rangle & = & p_l(T_c) =  p_h(T_c) \\
\langle \rho \rangle & = & f_l \rho_l(T_c) + (1-f_l) \rho_h(T_c) 
\label{eq:7}\\
\langle s \rangle & = & f_l s_l(T_c) + (1-f_l) s_h(T_c)\quad ,
\label{eq:8}
\end{eqnarray}
where $f_l$ is the fraction of volume existing in the low-energy phase.
During the phase transition the pressure and temperature stay constant and the evolution of average energy- and entropy- density is governed by the
evolution of the volume fraction $f_l$. It is therefore clear that over sufficiently
large regions the effective speed of sound is zero
\begin{equation}
c_S^{eff} = \sqrt{\biggl(\frac{\partial \langle p \rangle}{\partial \langle\rho \rangle}\biggr)_S}\approx 0
\end{equation}
These arguments of course only apply in thermodynamic equilibrium, and not for
super -cooled or -heated phase transitions. Using the constancy for entropy
during adiabatic evolution (i.e. $s a^3 = {\rm constant}$) as well as
Eqs.~\ref{eq:4},\ref{eq:5},\ref{eq:7}, and \ref{eq:8} one may derive the
evolution of average energy density with scale factor $a$ during a first-order phase transition
\begin{equation}
\langle \rho \rangle (a) = \biggl(\frac{a_0}{a}\biggr)^3[\rho_h + \frac{1}{3}\rho_l] -  \frac{1}{3}\rho_l \, .
\end{equation}
for $a_b < a < a_e$ with
$a_b$ and $a_e$ the scale factor at the beginning and the end of the phase transition, respectively. It is interesting to note that it
is more akin of a dust like phase (i.e. $\rho\sim 1/a^3$) than a radiation
dominated phase (i.e. $\rho\sim 1/a^4$). During dust like phases density
perturbations can grow unhindered by pressure forces. We therefore expect PBH formation during first-order phase transitions to be faciltated due to the absence of pressure forces.

\begin{figure}[htbp]
\centering
\includegraphics[width=0.52\textwidth]{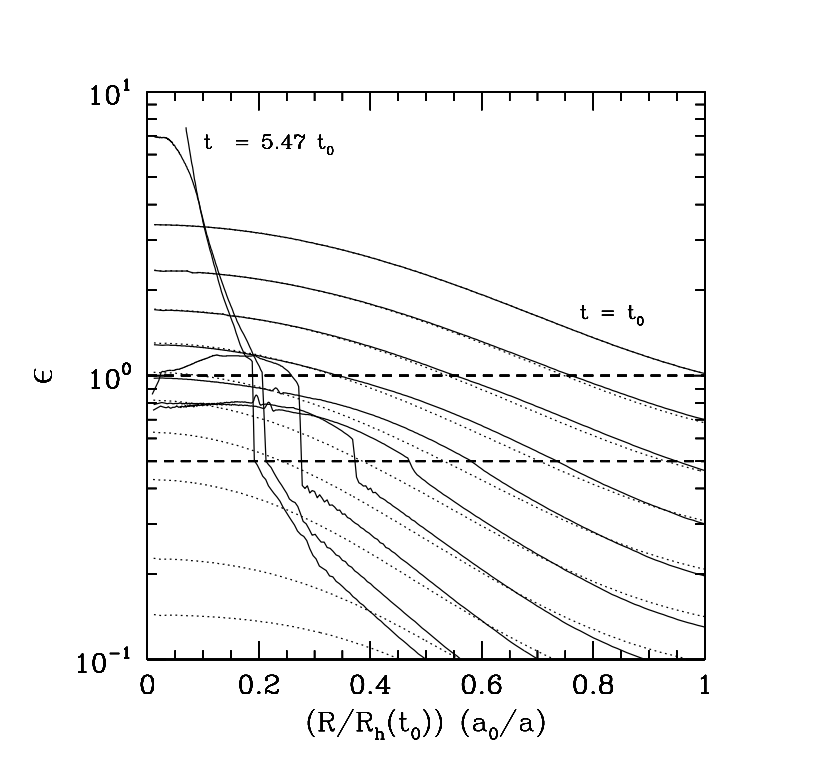}
\caption{The evolution of energy density as a function of radial
coordinate for a density fluctuation generated during inflation and re-entering the horizon after inflation. Results
for two scenarios are shown. (a) a first-order phase transition in the
density range $\rho = 0.5-1$ (solid) and (b) a radiation dominated Universe
without phase transition (dotted). Figure taken from \cite{Jedamzik:1999am}.
}
\label{fig:first_order}
\end{figure}

The PBH formation process during a short first-order phase transition has
been numerically investigated in \cite{Jedamzik:1999am} This general relativistic hydrodynamic
simulation of the evolution of a spherical symmetric overdensity shows the formation
of a PBH. In Fig.~\ref{fig:first_order} the evolution of the density profile
as a function of radial coordinate is shown. A first-order phase transition
with $c_s^{eff}\approx 0$ has been assumed in the energy density range 
$\rho = 0.5-1$, as indicated by the dashed lines, and $c_s = 1/\sqrt{3}$ 
for $\rho > 1$ and $\rho < 0.5$ has been adopted. Note that only the inner
region of the simulation is shown. It  is seen that at late times when the
fluctuation is collapsing any matter which reaches $\rho_l(T_c)$ from below
during the collapse, will go through the entire phase transition, until
$\rho \geq \rho_h(T_c)$ as there are no pressure forces to prevent this
collapse. This bias gives the fluctuation the extra kick to form a PBH.
In comparison, the dotted lines show the evolution of the density profile
in the absence of a first-order phase transition, where for the same initial
fluctuation no PBH is formed.

\subsection{QCD phase transition}

\begin{figure}[htbp]
\centering
\includegraphics[width=0.52\textwidth]{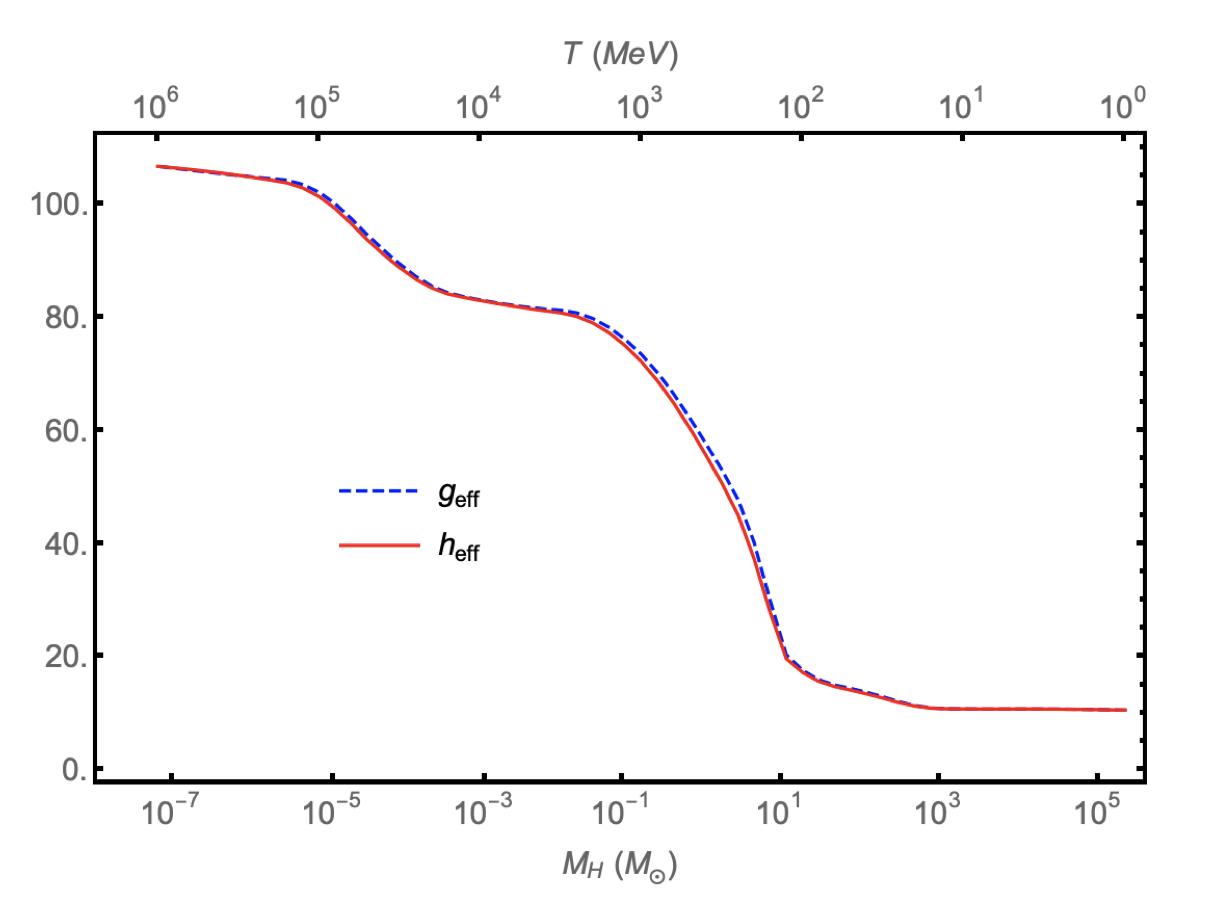}
\caption{The statistical weights for energy density $g_{\rm eff} = 
30\rho/(\pi^2 T^4)$ and entropy density $h_{\rm eff} = 
45 s/(2\pi^2 T^3)$ as a function of cosmic temperature $T$ and horizon 
mass $M_h(T)$.  
Kindly provided by the authors of \cite{Juan:2022mir}.
}
\label{fig:dof}
\end{figure}

\begin{figure}[htbp]
\centering
\includegraphics[width=0.52\textwidth]{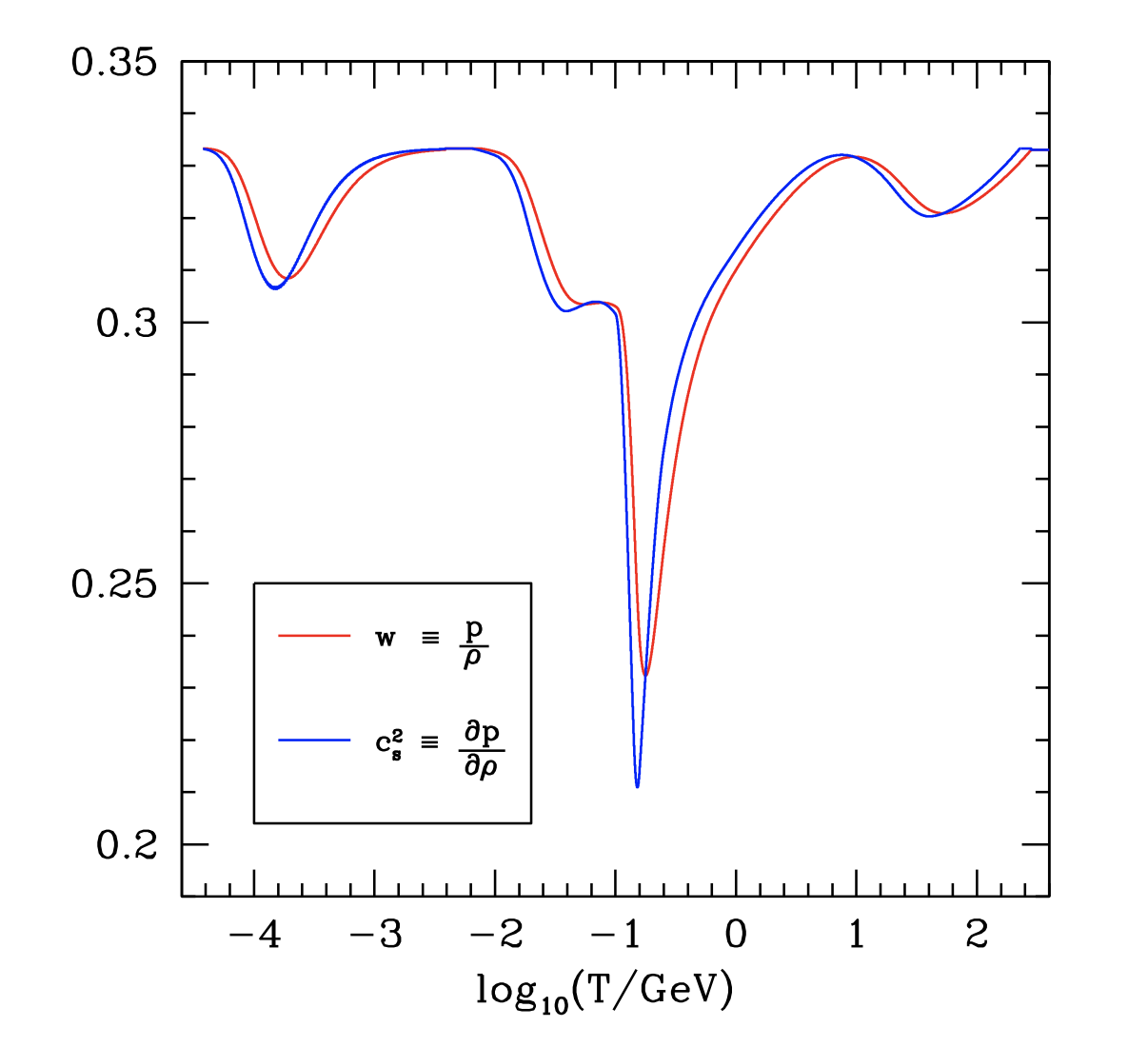}
\caption{The speed of sound squared $c_s^2$ and $w=p/\rho$ in the
early Universe for a wide range of temperatures. Taken from \cite{Musco:2023dak}.
}
\label{fig:speed_of_sound}
\end{figure}

In the seventies when nuclear physicists discovered a large number of
nuclear resonances, it was speculated that the associated softening of the equation of state could lead to an enhancement of PBH formation \cite{Chapline:1975tn}. With the
discovery of quantum chromo-dynamics it became clear that nuclear resonances
were not the fundamental particles, but were made up of quarks and gluons.
For a long time it was thought that the necessary transition from hadronic
matter (mesons and baryons) to their constituents (quarks and gluons) 
at $T\sim 100\,$MeV could be a first-order phase transition. It was speculated
that the softening of the equation of state could lead to an enhancement of
solar mass PBHs and potentially explain the MACHO signal of some compact dark
matter in the Milky Way halo \cite{Jedamzik:1996mr,Jedamzik:1998hc}.
\footnote{The question if there are some
massive compact Milky Way halo objects is still under debate, as the EROS
\cite{Tisserand:2006zx}
and MACHO \cite{Alcock:2000ph} collaborations find different results. It can, however, be
excluded that the Milky way halo is entirely made up of solar mass PBHs, as
they would violate microlensing constraints. This conclusion  can not be
circumvented when the PBHs are members of clusters \cite{Petac:2022rio}.}
Detailed lattice gauge simulations in this millenium 
\cite{Bhattacharya:2014ara,Borsanyi:2016ksw} seem to have established
that the QCD transition is no phase transition but a cross-over. There is nevertheless a reduction of speed of sound during the crossover. 
Fig.~\ref{fig:dof} shows the relativisitic degrees of freedom $g$ for cosmic
temperatures which include the electroweak- and QCD- phase transitions. It also
associates the horizon, i.e. the approximated PBH mass, with cosmic temperature. It is seen that during the QCD transition $g$ drops from $g\approx 80$ to $g\approx 11$.
In 
Fig.~\ref{fig:speed_of_sound} the square of the speed of sound $c_s^2$ and 
$w = p/\rho$ are shown for a wide range of temperatures between $T\approx 25\,$
keV and $T\approx 400\,$GeV. It is seen that, though the reduction
of $c_s^2$ from a value of $1/3$ is not as drastic as for a first-order phase transition where $c_s^2\approx 0$, that there are several epochs where
$c_s^2$ is reduced. Most notable is the reduction during the QCD transition
at $T\approx 100\,$MeV. It is relatively large due to the large number of
degrees of freedom involved. Smaller reductions occur during the electroweak
transition at $T\approx 100\,$GeV, the pion annihilation epoch at $T\approx 50\,$GeV, and the $e^+e^-$ annihilation epoch at $T\approx 200\,$keV. During
epochs where the plasma contains particles with mass $M\sim T$, a generic
reduction in $c_s^2$ and $w$ occurs, as rest mass energy does not contribute to pressure. Only when $T$ has dropped signifantly below $M$, thereby provoking
the annihilation of massive particles with their anti-particles, will $c_s^2$ bounce back to its radiation dominated value. This is the case for the massive
gauge bosons $W$ and $Z$ during the electroweak transition, massive pions
shortly after the QCD transition and massive electrons and positrons during
Big Bang nucleosynthesis.

\begin{figure}[htbp]
\centering
\includegraphics[width=0.52\textwidth]{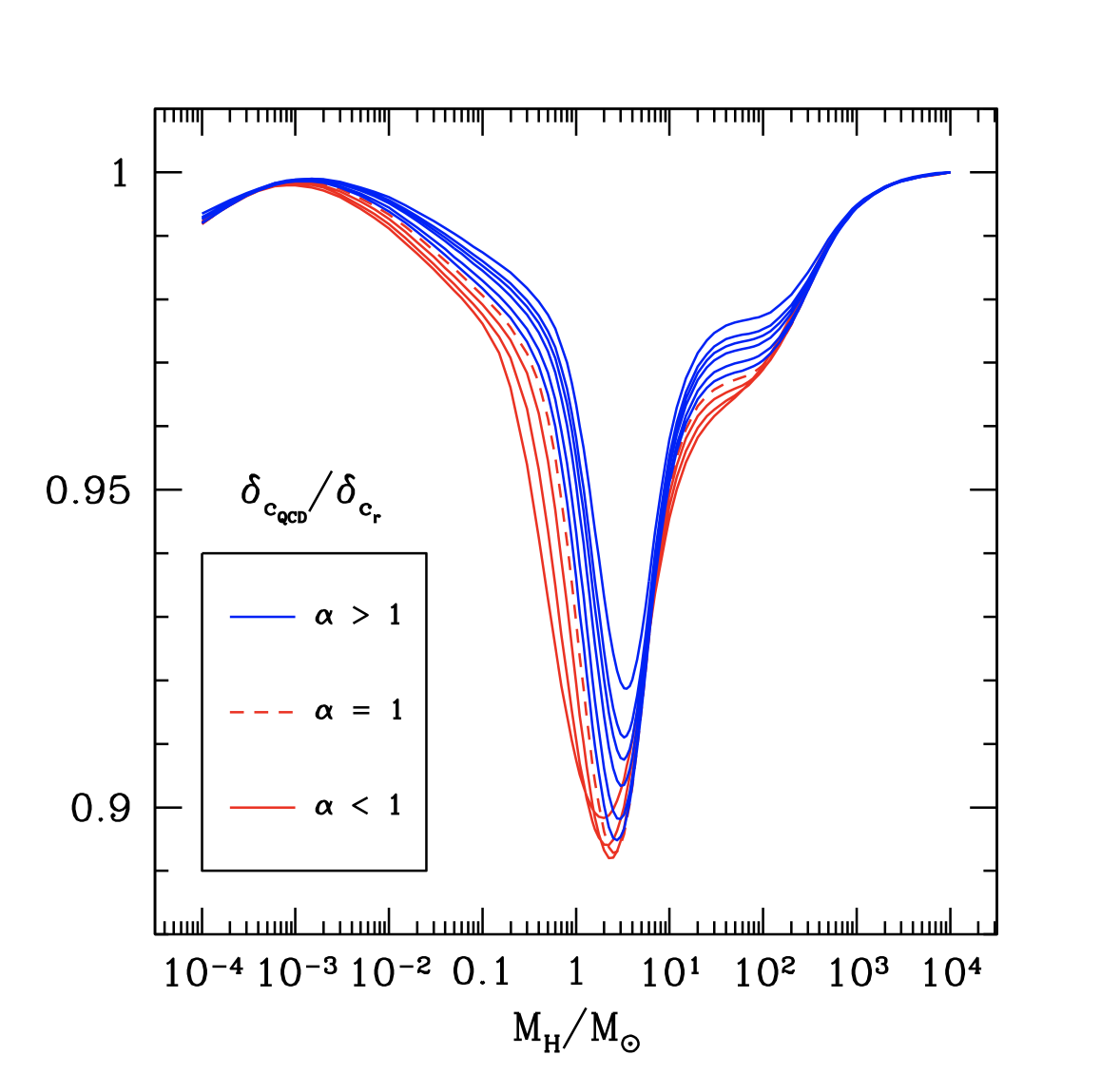}
\caption{The critical threshold for PBH formation during the QCD epoch normalized by the value of a radiation dominated Universe. 
The parameter $\alpha$ quantifies the shape of the fluctuation.  
Taken from \cite{Musco:2023dak}.
}
\label{fig:threshold}
\end{figure}

The detailed equation of state has been used to compute the density (or curvature) threshold $\delta_c$ at each epoch required for PBH formation.
Pre-existing fluctuations with $\delta\geq\delta_c$ form PBHs upon
horizon crossing, whereas fluctuations with $\delta <\delta_c$ are dispersed by pressure forces (cf. Chapter 5). These studies have been done by either approximate analytical collapse models 
\cite{Byrnes:2018clq,Carr:2019kxo} or by full numerical simulation
\cite{Escriva:2022bwe,Musco:2023dak}. In Fig. \ref{fig:threshold} the PBH formation threshold $\delta_c$ relative to the threshold in a purely radiation dominated Universe is shown.
The underlying study \cite{Musco:2023dak} used fully general-relativistic hydrodyamics 
simulations of spherical fluctuations with different fluctuation profiles
(given by the value $\alpha$). Only the QCD epoch is shown, leading, however,
to a comparatively large range in horizon masses ($\sim$ PBH masses). It is
seen that the $\sim 30\%$ reduction in $c_s^2$ leads to a $\sim 10\%$ 
reduction in threshold $\delta_c$. This study finds that for a wide range of
curvature fluctuations, the apparent horizon aways occurs first in conditions
when matter is at the minimum $c_s^2$. We have observed similar results for 
first order phase transitions in the previous section. It is also observed
that critcal scaling phenomena (Chapter 5) hold even in the QCD case with a varying equation of state, albeit with a different exponent than that
in a plasma with a purely radiation dominated Universe.

\begin{figure}[htbp]
\centering
\includegraphics[width=0.82\textwidth]{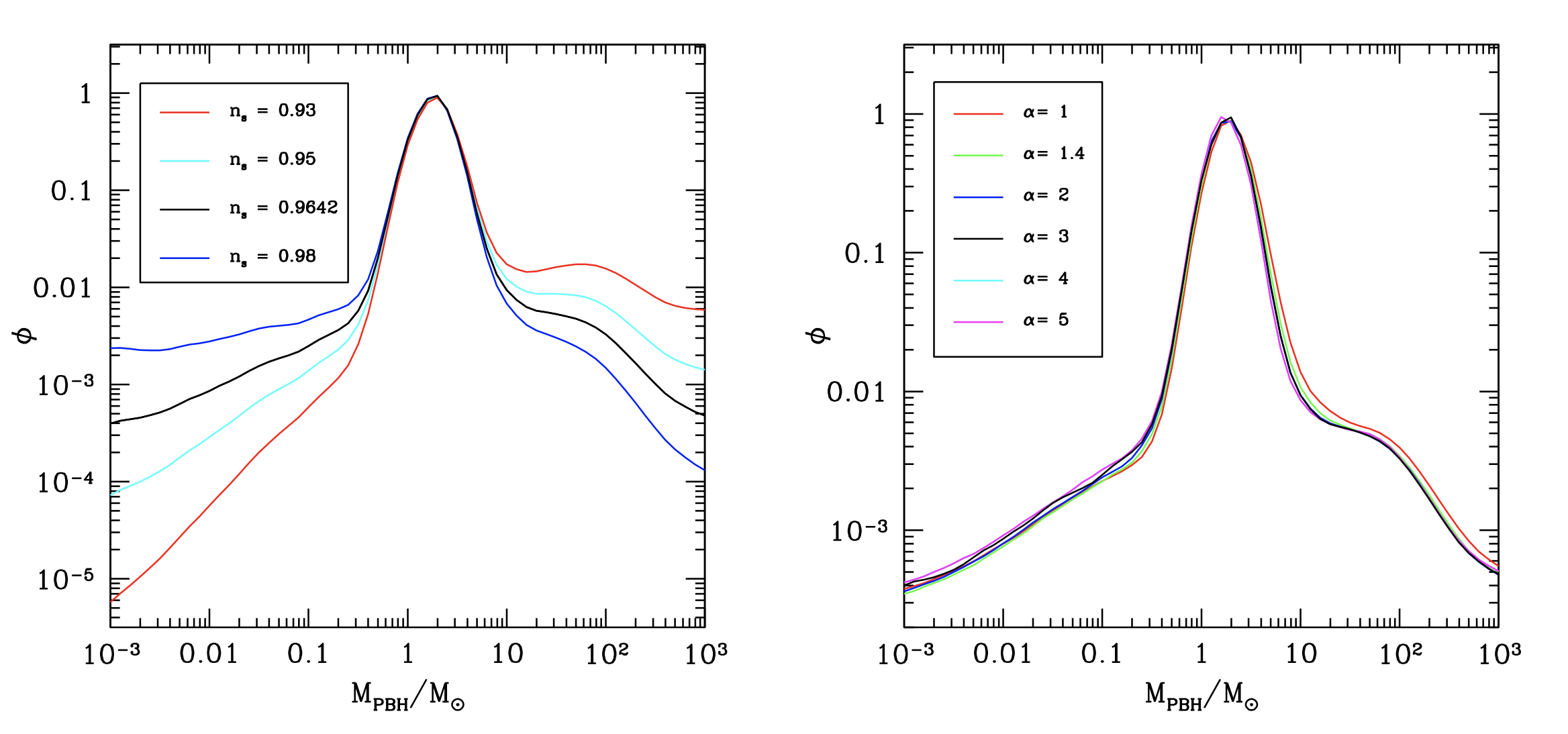}
\caption{The QCD PBH mass function $\Phi (M_{\rm PBH})$. Left panel:
The mass function for varying spectral index $n_s$ and $\alpha = 3$. Right panel: The mass function for varying $\alpha$ and $n_s = 0.9642$.
Taken from \cite{Musco:2023dak}.
}
\label{fig:mass_function}
\end{figure}

The results shown in Fig.~\ref{fig:threshold} may be used to compute
the QCD PBH mass function as a function of spectral index and fluctuation
shape $\alpha$. The calculations take full account of critical phenomena
and apply peaks theory (Chapter 7) for the computation of the PBH mass function.
Results are shown in Fig.~\ref{fig:mass_function} with the assumption of pre-existing Gaussian fluctuations. It is seen that the 
resulting peak in the mass function, which is at $M\approx 1.9M_{\odot}$,
is surprisingly independent of the fluctuation shape parameter $\alpha$ and spectral 
index $n_s$ of the underlying fluctuations. Of course the latter holds 
only when the fluctuations are approximately scale-invariant. For very blue
spectra $n_s > 1$ PBH formation is dominated at the smallest scales, and
vice versa for very red spectra $n_s < 1$. In such cases the QCD peak would not be very important. But for approximately scale-invariant spectra, PBH 
formation on the QCD scale $M_{\rm pbh} \approx 1.9 M_{\odot}$ is a factor
thousand more probable than on other scales. It is also noted that the
pion annihilation epoch leads to a small enhancement on mass scales
$M_{\rm pbh} > 10 M_{\odot}$. Approximately one per cent of all QCD PBHs
are thus massive. If some fraction of the observed LIGO/Virgo/Kagra 
$M > M_{\odot}$ events are PBHs ~\cite{Juan:2022mir,Franciolini:2022tfm}, then gravitational wave detectors have only
observed the tip of the iceberg, as there would be a factor $\sim 100$ more
low mass PBHs. Of course, the obvious smoking gun for the existence of
PBH dark matter in our Universe would be the discovery of PBHs of 
sub-Chandrasekar mass. Such PBHs are formed in QCD enhanced PBH formation
scenarios in quite large adundance.

\subsection{The $e^+e^-$ Annihilation Epoch}

It has been noted \cite{Jedamzik:1996mr,Carr:2019kxo}, 
that the decrease of $w$ and $c_s^2$
during the $e^+e^-$ annihilation epoch could lead to an enhancement of the abundance of PBHs on the \mbox{$M_{\rm pbh}\sim 10^5M_{\odot}$} scale. However this may not necessarily be the case, as the cosmic $e^+e^-$ annihilation epoch is quite different from the QCD epoch and the electroweak epoch. At temperatures $T\approx 1-2\,$MeV, shortly before the $e^+e^-$ annihilation, with a maximum reduction of $w$ at $T\approx 200\,$keV, neutrinos decouple from the
Universe. In particular, since their interactions with the rest of the plasma
freeze out, neutrinos can free-stream out of overdense regions, effectively
reducing the overdensity. One may schematically write for the overdensity
\begin{equation}
\frac{\delta\rho}{\rho} = \frac{\sum\limits_i \delta\rho_i}{\sum\limits_i \langle\rho_i\rangle} \, ,
\end{equation}
where brackets denote cosmic average and the index $i$ runs over particle species. For adiabatic perturbations one has 
$\delta\rho_i = K\langle\rho_i\rangle$ with $K$ a quantity independent of
species, such that $\delta\rho/\rho = K$. However, the free-streaming of the neutrinos destroys the adiabaticity of the perturbation since 
$\delta\rho_{\nu}\approx 0$.
Having $\langle\rho_{\nu}\rangle/\langle\rho_{\rm tot}\rangle = g_{\nu}/g_{\rm tot} = 5.25/10.75 \approx 0.5$ one may estimate that the original perturbation
has only approximately half of the overdensity after neutrino free-streaming.
On the other hand, the critical threshold for PBH formation will not reduce
by as much as a factor of two due to the $e^+e^-$ equation of state, such that
the argument would imply that formation of PBHs during any epoch after neutrino coupling is highly suppressed for scale invariant perturbation spectra. Seen from the
opposite point of view, even if neutrinos are initially homogeneous, the
density perturbation in photons and $e^+e^-$ would gravitationally attract
neutrinos. Since those neutrinos do not even exert pressure, a further 
reduction of w to $w\sim 0.33/2$ would occur, favoring PBH formation. 
Ref. \cite{Musco:2023dak}
argues that the first argument may be dominant, suppression of PBH formation.
However, only a dedicated simulation of PBH formation with a fluid and free-streaming component could definitely answer this question. The formation
of PBHs during the $e^+e^-$ annihilation is therefore an open question.

\bibliography{pbhchapter}



\end{document}